\documentclass[apj]{emulateapj}
\usepackage{amsmath}
\usepackage{graphicx}
\usepackage{enumitem}
\usepackage{multirow}
\usepackage{color}
\usepackage[dvipsnames]{xcolor}
\usepackage[T1]{fontenc}
\usepackage{ae,aecompl}
\usepackage{amssymb}	% Extra maths symbols
\usepackage{hyperref}
\hypersetup{colorlinks, citecolor=blue, linkcolor=blue}

\usepackage{natbib}
\def\aj{AJ}% Astronomical Journal
\def\apj{ApJ}% Astrophysical Journal
\def\apjl{ApJ}% Astrophysical Journal, Letters
\def\apjs{ApJS}% Astrophysical Journal, Supplement
\def\aap{A\&A}% Astronomy and Astrophysics
\def\mnras{MNRAS}% Monthly Notices of the RAS
\def\nat{Nature}% Nature
\def\araa{ARA\&A}

\def\pasj{PASJ}
\begin{document}

\title{The SAMI Galaxy Survey: Stellar Populations of Passive Spiral Galaxies in different environments}

\author{Mina Pak\altaffilmark{1,2},
Sree Oh\altaffilmark{3,4}, 
Joon Hyeop Lee\altaffilmark{1,2}, 
Nicholas Scott\altaffilmark{4,5},
Rory Smith\altaffilmark{1},
Jesse van de Sande\altaffilmark{4,5},
Scott M. Croom\altaffilmark{4,5},
Francesco D'Eugenio\altaffilmark{6},
Kenji Bekki\altaffilmark{7},
Sarah Brough\altaffilmark{4,8},
Caroline Foster\altaffilmark{4,5},
Tania M. Barone\altaffilmark{3,4,5}, 
Katarina Kraljic\altaffilmark{9}, 
Hyunjin Jeong\altaffilmark{1}, 
Joss Bland-Hawthorn\altaffilmark{5},
Julia J. Bryant\altaffilmark{4,5,10},
Michael Goodwin\altaffilmark{4,5,11},
Jon Lawrence\altaffilmark{12},
Matt S. Owers\altaffilmark{13,14},
Samuel N. Richards\altaffilmark{15}
}
\affil{\altaffilmark{1}Korea Astronomy and Space Science Institute (KASI), 776 Daedukdae-ro, Yuseong-gu, Daejeon  34055, Republic of Korea}
\affil{\altaffilmark{2}University of Science and Technology, Korea (UST), 217 Gajeong-ro Yuseong-gu, Daejeon 34113, Republic of Korea}
\affil{\altaffilmark{3}Research School of Astronomy and Astrophysics, Australian National University, Canberra, ACT 2611, Australia}
\affil{\altaffilmark{4}ARC Centre of Excellence for All Sky Astrophysics in 3 Dimensions (ASTRO 3D), Australia}
\affil{\altaffilmark{5}Sydney Institute for Astronomy (SIfA), School of Physics, University of Sydney, NSW 2006, Australia}
\affil{\altaffilmark{6}Sterrenkundig Observatorium, Universiteit Gent, Krijgslaan 281 S9, B-9000 Gent, Belgium}  
\affil{\altaffilmark{7}ICRAR M468, The University of Western Australia, 35 Stirling Hwy, Crawley, Western Australia 6009, Australia}
\affil{\altaffilmark{8}School of Physics, University of New South Wales, NSW 2052, Australia}
\affil{\altaffilmark{9}Institute for Astronomy, Royal Observatory, Edinburgh EH9 3HJ, UK}
\affil{\altaffilmark{10}Australian Astronomical Optics, AAO-USydney, School of Physics, University of Sydney, NSW 2006, Australia}
\affil{\altaffilmark{11}Australian Astronomical Optics – Macquarie, 105 Delhi Rd, North Ryde, NSW 2113, Australia}
\affil{\altaffilmark{12}Australian Astronomical Optics - Macquarie, Macquarie University, NSW 2109, Australia}
\affil{\altaffilmark{13}Department of Physics and Astronomy, Macquarie University, NSW 2109, Australia}
\affil{\altaffilmark{14}Astronomy, Astrophysics and Astrophotonics Research Centre, Macquarie University, Sydney, NSW 2109, Australia}
\affil{\altaffilmark{15}SOFIA Science Center, USRA, NASA Ames Research Center, Building N232, M/S 232-12, P.O. Box 1, Moffett Field, CA 94035-0001, USA}

\email{minapak@kasi.re.kr}

\begin{abstract}
We investigate the stellar populations of passive spiral galaxies as a function of mass and environment, using integral field spectroscopy data from the Sydney-AAO Multi-object Integral field spectrograph Galaxy Survey. Our sample consists of $52$ cluster passive spirals and $18$ group/field passive spirals, as well as a set of S0s used as a control sample. The age and [Z/H] estimated by measuring Lick absorption line strength indices both at the center and within $1R_{\rm e}$ do not show a significant difference between the cluster and the field/group passive spirals. However, the field/group passive spirals with log(M$_\star$/M$_\odot)\gtrsim10.5$ show decreasing [$\alpha$/Fe] along with stellar mass, which is $\sim0.1$ dex smaller than that of the cluster passive spirals. We also compare the stellar populations of passive spirals with S0s. In the clusters, we find that passive spirals show slightly younger age and lower [$\alpha$/Fe] than the S0s over the whole mass range. In the field/group, stellar populations show a similar trend between passive spirals and S0s. In particular, [$\alpha$/Fe] of the field/group S0s tend to be flattening with increasing mass above log(M$_\star$/M$_\odot)\gtrsim10.5$, similar to the field/group passive spirals. We relate the age and [$\alpha$/Fe] of passive spirals to their mean infall time in phase-space; we find a positive correlation, in agreement with the prediction of numerical simulations. We discuss the environmental processes that can explain the observed trends. The results lead us to conclude that the formation of the passive spirals and their transformation into S0s may significantly depend on their environments. 

\end{abstract}
%\keywords{galaxies: evolution}

%%%%%%%%%%%%%%%%%%%%%%
\section{Introduction}
%%%%%%%%%%%%%%%%%%%%%%
% 1st
Stellar populations of nearby galaxies allow us to better understand the formation histories of galaxies. Red galaxies, which are generally non star-forming, are more likely to be spheroid dominated, whereas blue galaxies, young and star-forming, are more likely to be disk dominated, and to have spiral arms (e.g. \citealt{Hum36}; \citealt{Fio99};  \citealt{Ber00}; \citealt{Mig09}). There is however a population of galaxies that bridges both classes: passive spirals are red and quiescent, yet have prominent spiral patterns typical of star-forming galaxies (e.g. \citealt{van76}; \citealt{Wil80}; \citealt{Bot80}; \citealt{Phi88}; \citealt{Cay94}). After the discovery of `anemic' spiral galaxies in the Virgo cluster by \citet{van76} and a significant number of passive spirals with apparently no/weak sign of ongoing star formation have been found in all environments, both at low and high redshift. The formation and evolution of these galaxies have received considerable attention observationally and theoretically (e.g. \citealt{Cou98}; \citealt{Dre99}; \citealt{Pog99}; \citealt{Bek02}; \citealt{Got03}; \citealt{Yam04}; \citealt{Mor06}; \citealt{Cor09}; \citealt{Hug09}; \citealt{Lee08}; \citealt{Mah09}; \citealt{Wol09}; \citealt{Mas10}; \citealt{Fra16}; \citealt{Fra18}; \citealt{Pak19}). Despite this large number of studies, several questions still remain debated, such as (1) the origin(s) of passive spirals, (2) physical processes responsible for shutting down star formation without destroying their spiral structures, and (3) evolutionary connections with S0s.

% \citep[e.g.][]{Cou98; Dre99; Pog99; Bek02; Got03; Yam04; Mor06; Cor09; Hug09; Lee08; Mah09; Wol09; Mas10; Fra16; Fra18}
% 2nd
 The existence of passive spirals may be the evidence for the morphological transformation of spirals into S0s. The numerical simulations from \citet{Bek02} show how cluster environmental quenching processes can transform spirals into S0s, passing through an intermediate passive spiral phase. The spiral arm structures fade over several Gyrs, after the gas is stripped. However, they also claim that the passive spirals are found anywhere from galaxies in isolation to the centers of clusters, and hence that no single mechanism can completely explain their origin. 

% 3rd
\citet{Got03} suggested that the formation of passive spirals in the intermediate to high-density environments is closely related with cluster environmental effects: ram pressure stripping \citep{Gun72}, harassment \citep{Moo99}, thermal evaporation \citep{Cow77}, strangulation (\citealt{Lar80}), galaxy-galaxy interactions including major \citep{Too72} and minor \citep{Wal96} mergers and tidal interactions.

% 4th
Secular evolution possibly caused by a bar is another possible mechanism for quenching spiral galaxies (\citealt{Kor04}; \citealt{Ath13}), which may convert them into S0s. Bars are common structures in disk galaxies in the local universe. Recent work on the bars of nearby disk galaxies suggests that the bar fraction is up to $\sim 50\%$ in the optical bands (\citealt{Mar07}; \citealt{Ree07}; \citealt{Bar08}). The fraction rises to $\sim 70\%$ in near-infrared studies (\citealt{Kna00}; \citealt{Men07}). While all of these devoted studies have drawn attention to passive spirals, it is still under debate which is the dominant pathway of evolution from the passive spirals to S0s.

% 5th
To find the evolutionary connections between passive spirals and S0s in the perspective of stellar populations, \citet{Pak19} investigated nine passive spirals using the Calar Alto Legacy Integral Field Area (CALIFA) survey (\citealt{San12}; \citealt{Hus13}). When comparing stellar populations (see figures 5 and 7 in \citealt{Pak19}), the S0s were found to have a wider range of age, metallicity and $\alpha$-abundance than the passive spirals at all radii out to $2 R_{\rm e}$. The stellar populations of passive spirals were fully encompassed within the spread of the S0 stellar populations, although the distributions of the passive spirals appeard skewed toward younger ages, higher metallicities, and lower $\alpha$-abundances. However, the environmental dependence was weakly constrained because the number of passive spirals was too small. Thus, investigating the environments of passive spirals using a larger sample may provide further constraints on their quenching mechanisms.

% 6th
Here, we investigate the spatially resolved stellar populations of passive spirals as a function of their environment using the Sydney-Australian Astronomical Observatory Multi-object Integral Field Spectrograph (SAMI; \citealt{Cro12}) Galaxy Survey \citep{Bry15} and the Galaxy and Mass Assembly (GAMA; \citealt{Dri11}; \citealt{Hop13}) survey. In this study, we aim to explore the environmental dependence of the formation and evolution of passive spirals and their evolutionary connections with S0s, and to provide further constraints on the quenching mechanisms.

% 7th
This paper is organized as follows. Section 2 describes the galaxy samples and the analysis methods. The stellar population properties of passive spirals depending on environments and the comparison with S0s are shown in Section 3. The formation and evolution of passive spirals is discussed in Section 4. Finally, in Section 5, we summarize our results and draw conclusions. Throughout the paper we adopt a standard $\Lambda$CDM cosmology with $\Omega_{m} = 0.3$, $\Omega_{\Lambda} = 0.7$, and $H_{0} = 70$ km s$^{-1}$ Mpc$^{-1}$. All broadband data are given in the AB photometric system.

%%%%%%%%%%%%%%%%%%%%%%%%%
\section{Data and Sample}
%%%%%%%%%%%%%%%%%%%%%%%%%
The SAMI Galaxy Survey is a spatially-resolved spectroscopic survey for over $3000$ galaxies in the redshift range of $0.004 <$ z $< 0.095$. The SAMI instrument \citep{Cro12} using $13$ fused fiber bundles (Hexabundles, \citealt{BH11}; \citealt{Bry14}) is mounted at the prime focus on the $3.9$m Anglo-Australian Telescope at Siding Spring Observatory in Australia. SAMI fibers are fed to the double-beam AAOmega spectrograph (\citealt{Sau04}; \citealt{Smi04}; \citealt{Sha06}). Details of the galaxy sample of SAMI Galaxy Survey are described in \citet{Bry15}, and additional details of the cluster sample are described in \citet{Owe17}. Besides galaxies from eight clusters, over $2000$ galaxies were targeted from the three $4^{\circ} \times 12^{\circ}$ GAMA regions centered on $9, 12$, and $15$h right ascension at dec $= 0$ (G09, G12, and G15 regions; \citealt{Dri11}). For all observed galaxies, the SAMI Galaxy Survey measured resolved optical properties such as star formation rate, age, metallicities, and stellar and ionized gas kinematics (\citealt{Van17}; \citealt{Sco18}). 

%%%%%%%%%%%%%%%%%%%%%%%%%%%%%
\subsection{Sample Selection}  % 2.1
%%%%%%%%%%%%%%%%%%%%%%%%%%%%%
%---------------- F1
\begin{figure*}
\includegraphics[width=18cm]{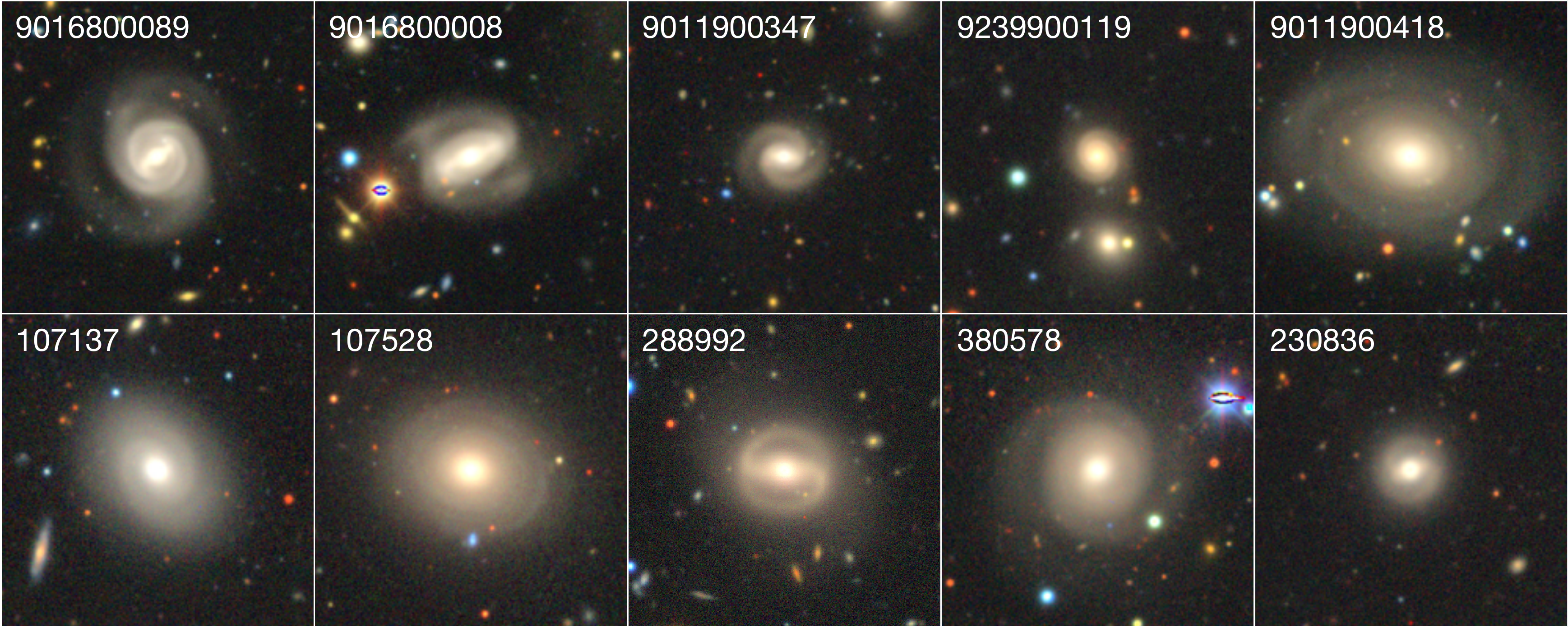}
\caption{The DECam Legacy Survey (DECaLS) images of the passive spiral galaxies in the SAMI-cluster (top row) and the SAMI-GAMA samples (bottom row). The image size is $100\arcsec \times 100\arcsec$. The ID of the SAMI galaxies is shown in the top corner of each galaxy.
\label{F1}}
\end{figure*}
%----------------
%---------------- F2
\begin{figure}
\includegraphics[width=8.5cm]{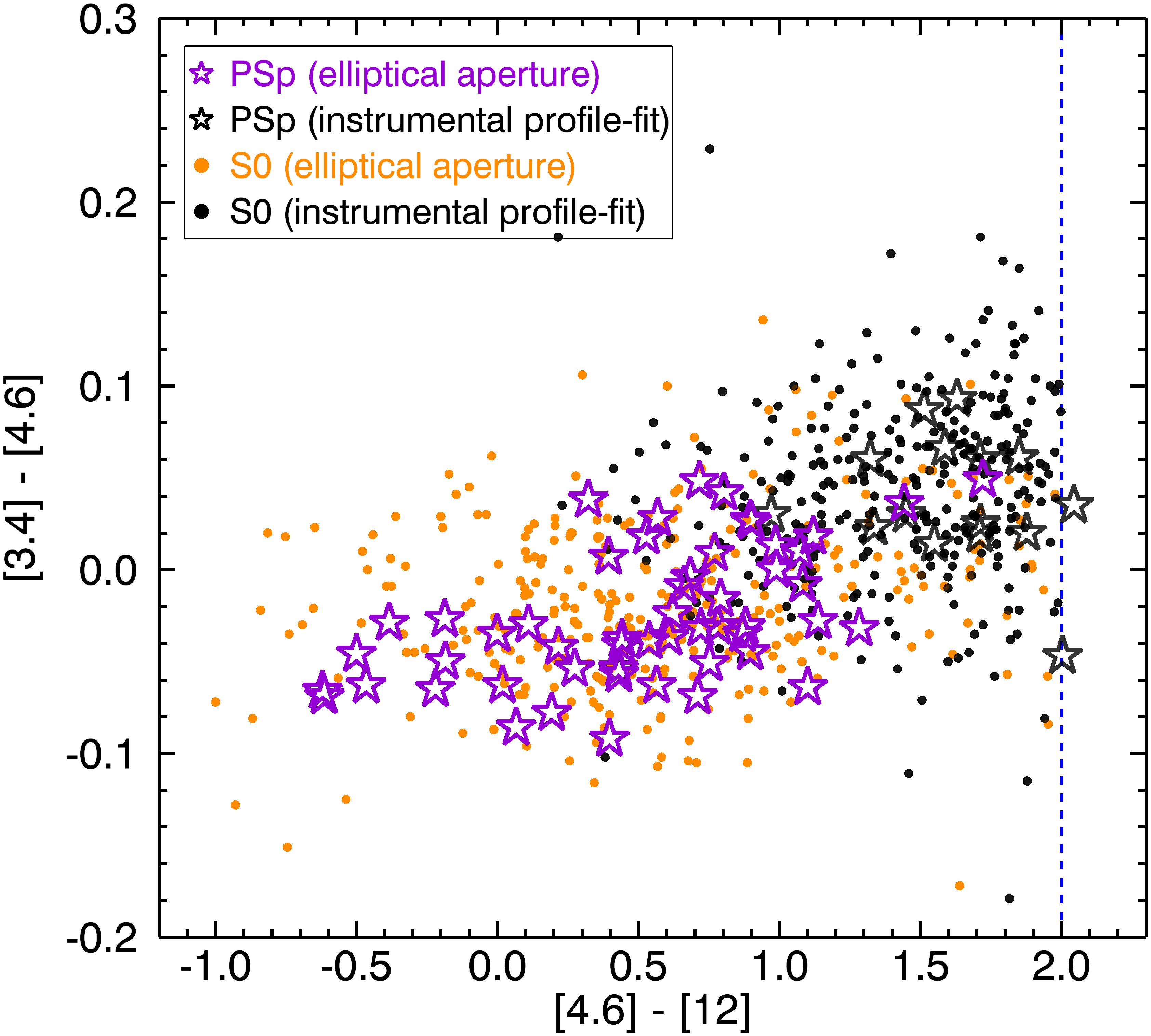}
\caption{The WISE color-color diagram for passive spiral (PSp) and S0s from the SAMI Galaxy Survey. Stars are the passive spirals and dots are the S0s. We use the magnitudes in the WISE catalog extracted using elliptical aperture photometry. When the elliptical-aperture magnitude is not available, we use the instrumental profile-fit photometry magnitude (black open stars for the passive spirals and black dots for the S0s). The dashed line distinguishes the quiescent galaxies (spheroidals) from intermediate disk galaxies \citep{Jar17}. 
\label{F2}}
\end{figure}
%----------------

%---------------- F31
\begin{figure}
\includegraphics[width=8.4cm]{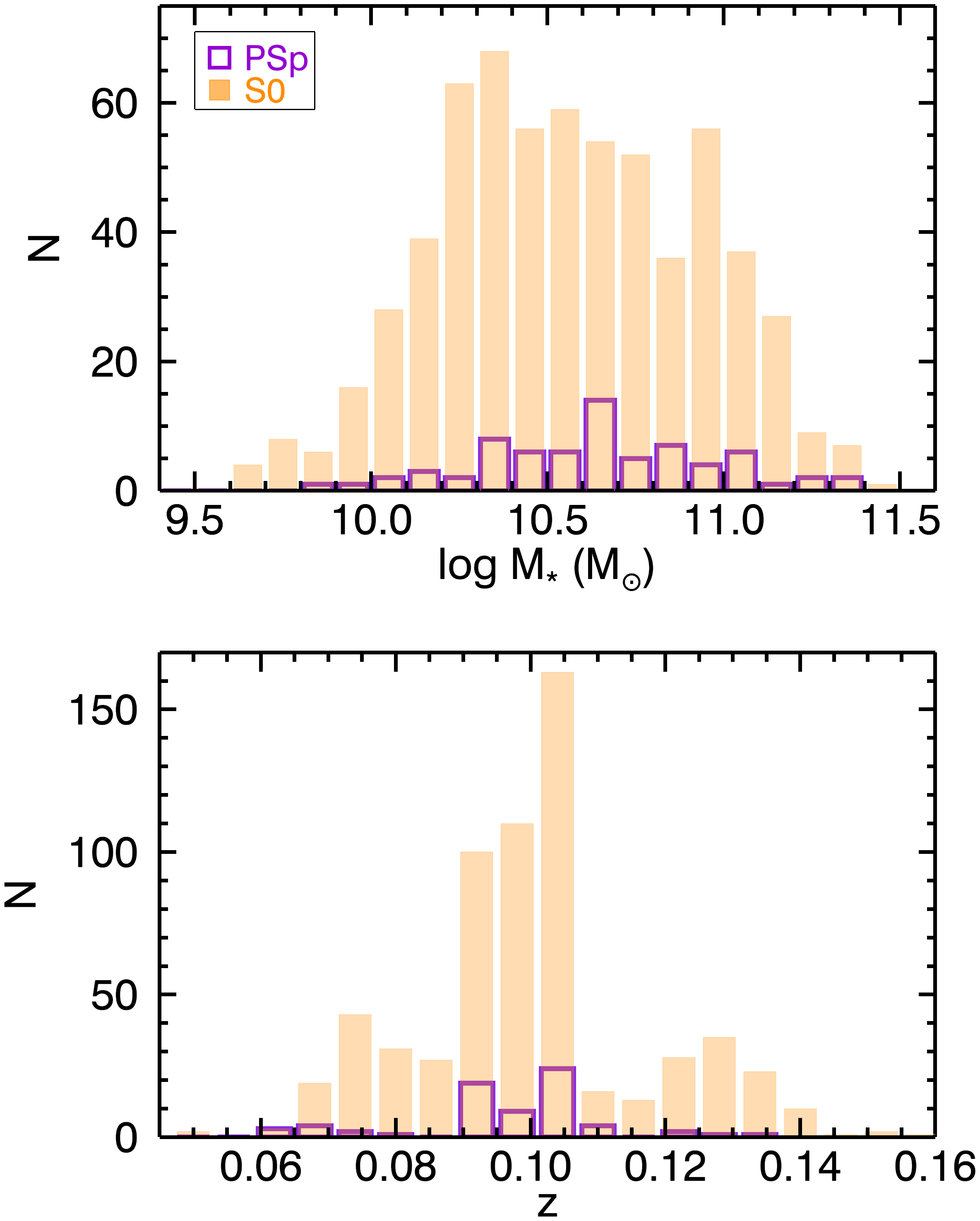}
\caption{The distributions of the stellar mass and redshift for the passive spirals (violet) and the S0s (orange shaded) in cluster and field/group environments.
\label{F31}}
\end{figure}
%----------------

%---------------- F32
\begin{figure}
\includegraphics[width=8.3cm]{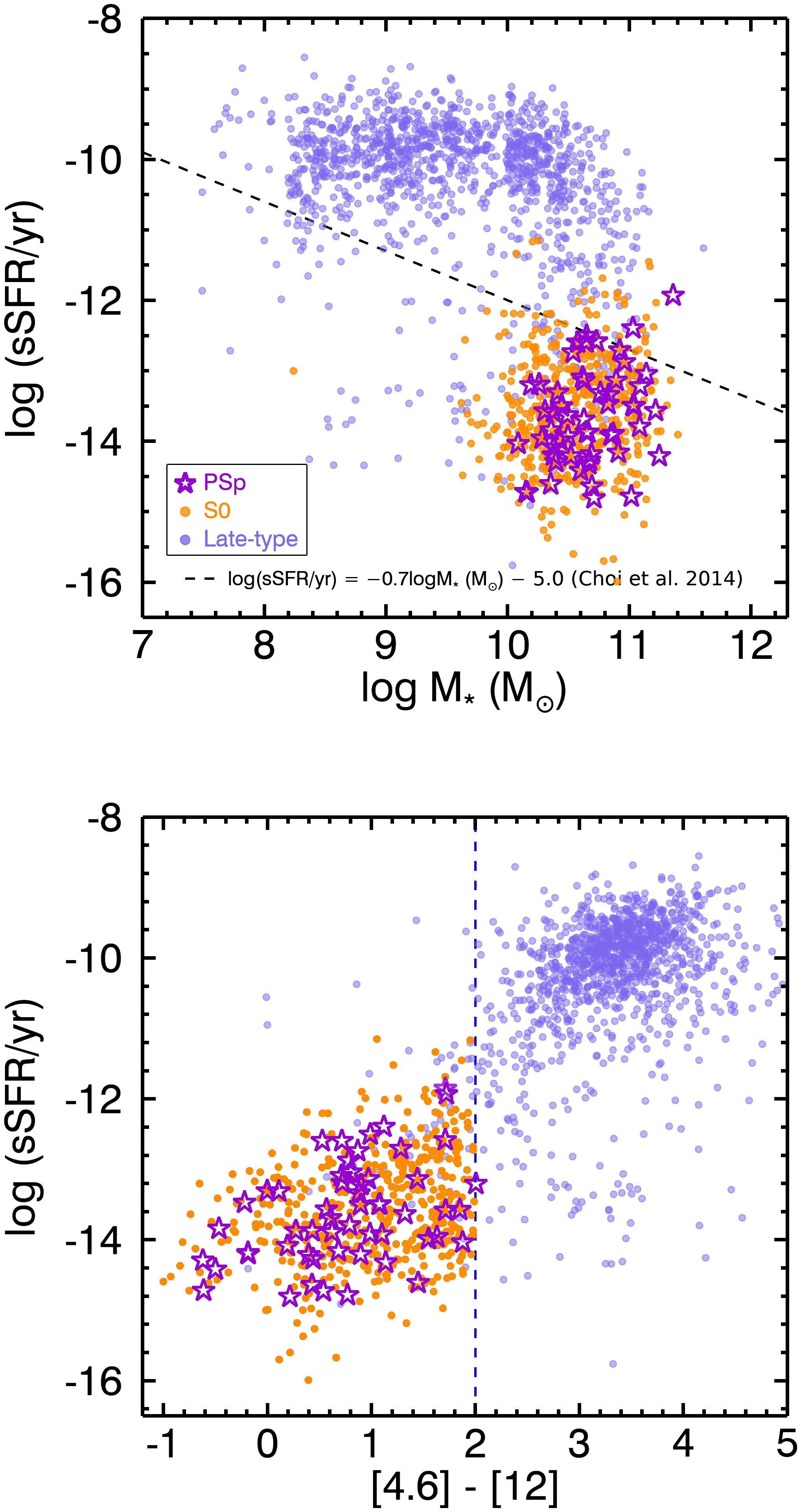}
\caption{Specific star formation rate \citep{Med18} as a function of stellar mass (top panel) and the WISE [4.6]-[12] color (bottom panel) for the passive spirals (stars) and the S0s (orange circles) drawn in Figure \ref{F2}. We also mark late-type galaxies (blue circles) from the SAMI Galaxy Survey data to infer that our sample selected by the WISE color-color diagram is well distinguished from star-forming late-type galaxies. The threshold of quiescent galaxies adopted from \citet{cho14} is shown as a dashed line in the top panel. The blue dashed line in the bottom panel is the same as Figure \ref{F2}.
\label{F32}}
\end{figure}
%----------------

%---------------------------------------- F4
\begin{figure*}
\includegraphics[width=18cm]{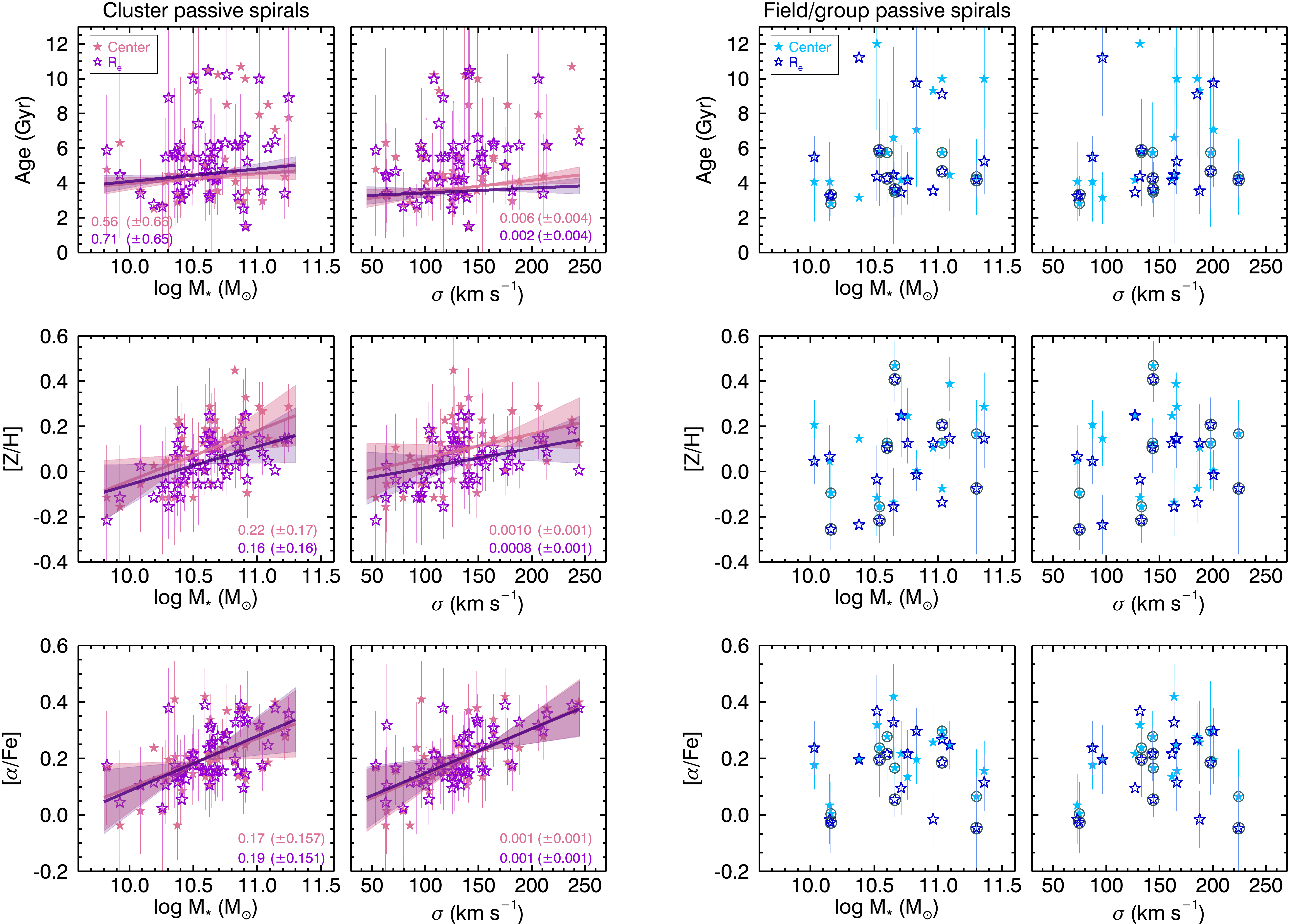}
\caption{Stellar mass and $\sigma_{\star}$ vs. age, [Z/H], and [$\alpha$/Fe] within $1.4$ arcsec (center) and $R_{\rm e}$ for the passive spirals (PSp) in the clusters (left) and the field/groups (right). Solid lines are linear-fits by weighted errors using the MPFIT function \citep{Mar09} in IDL library. In clusters, the slope ($\alpha$) of the fit is presented in the bottom of the panels and shaded areas show slope uncertainties. The six isolated passive spirals in field/group environments are enclosed with gray circles.
\label{F4}}
\end{figure*}
%---------------------------------------- 
%---------------------------------------- F5
\begin{figure*}
\includegraphics[width=18cm]{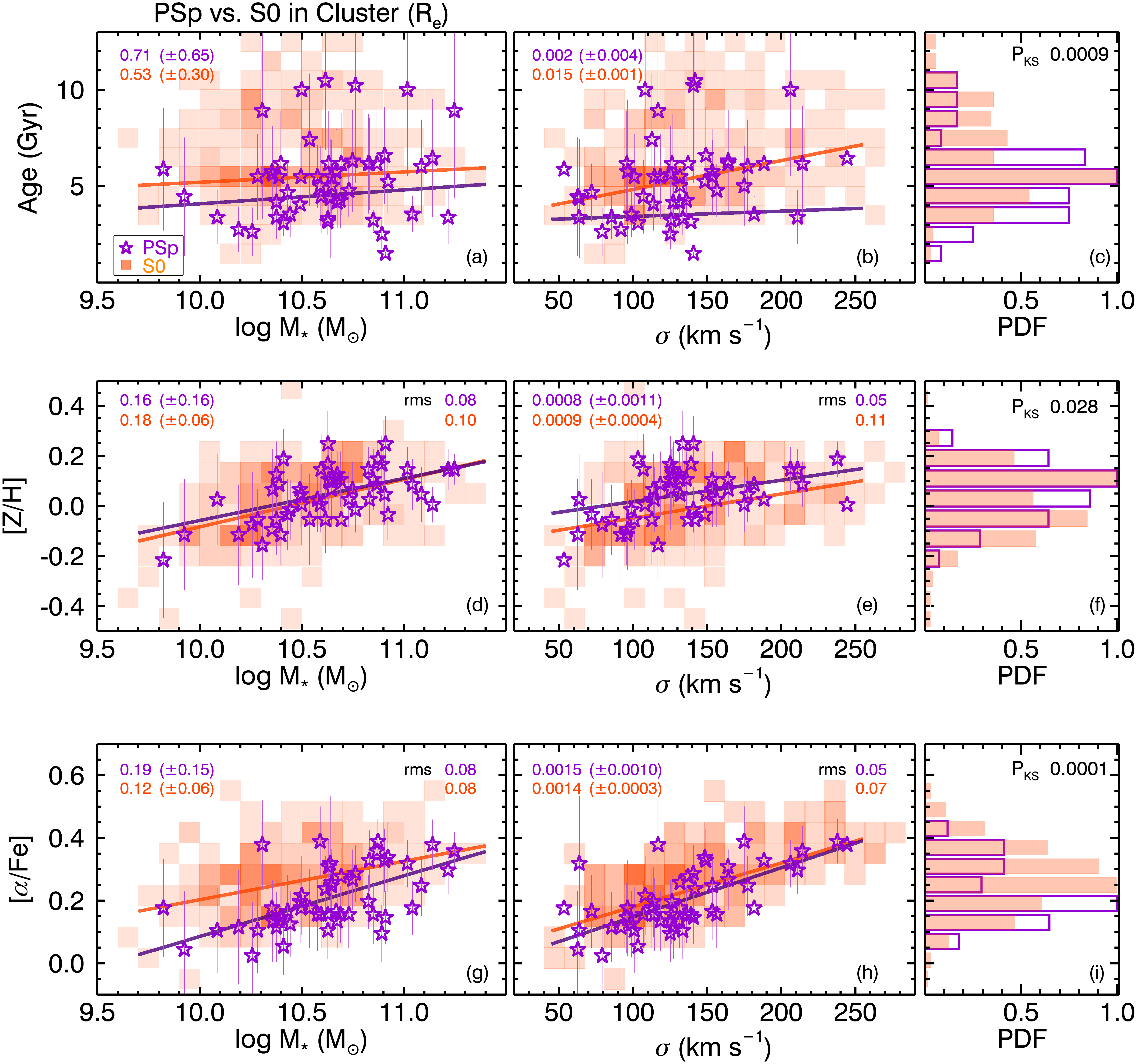}
\caption{Age (top), [Z/H] (middle), and [$\alpha$/Fe] (bottom) as a function of stellar mass and $\sigma_{\star}$ within $R_{\rm e}$ for the cluster sample, compared between the cluster passive spirals (stars) and the S0s (colorscale). Solid lines are linear-fits by weighted errors using the MPFIT function \citep{Mar09} in IDL library. The slope of the fit is presented in the top left corner of the panels. For passive spirals and S0s more massive than log (M$_\star$/M$_\odot) = 10.5$ ($\sigma = 150$ km s$^{-1}$), the rms scatter from each fit is shown in the top right corner of the panels. In each histogram, the probability of similarity in the distributions without consideration of stellar mass between the passive spirals and the S0s from a Kolmogorov-Smirnov test (P$_{KS}$) is given at the top right corner of each histogram. The violet and orange bins show passive spirals and S0s, respectively.
\label{F5}}
\end{figure*}
%----------------------------------------
%---------------------------------------- F6
\begin{figure*}
\includegraphics[width=18cm]{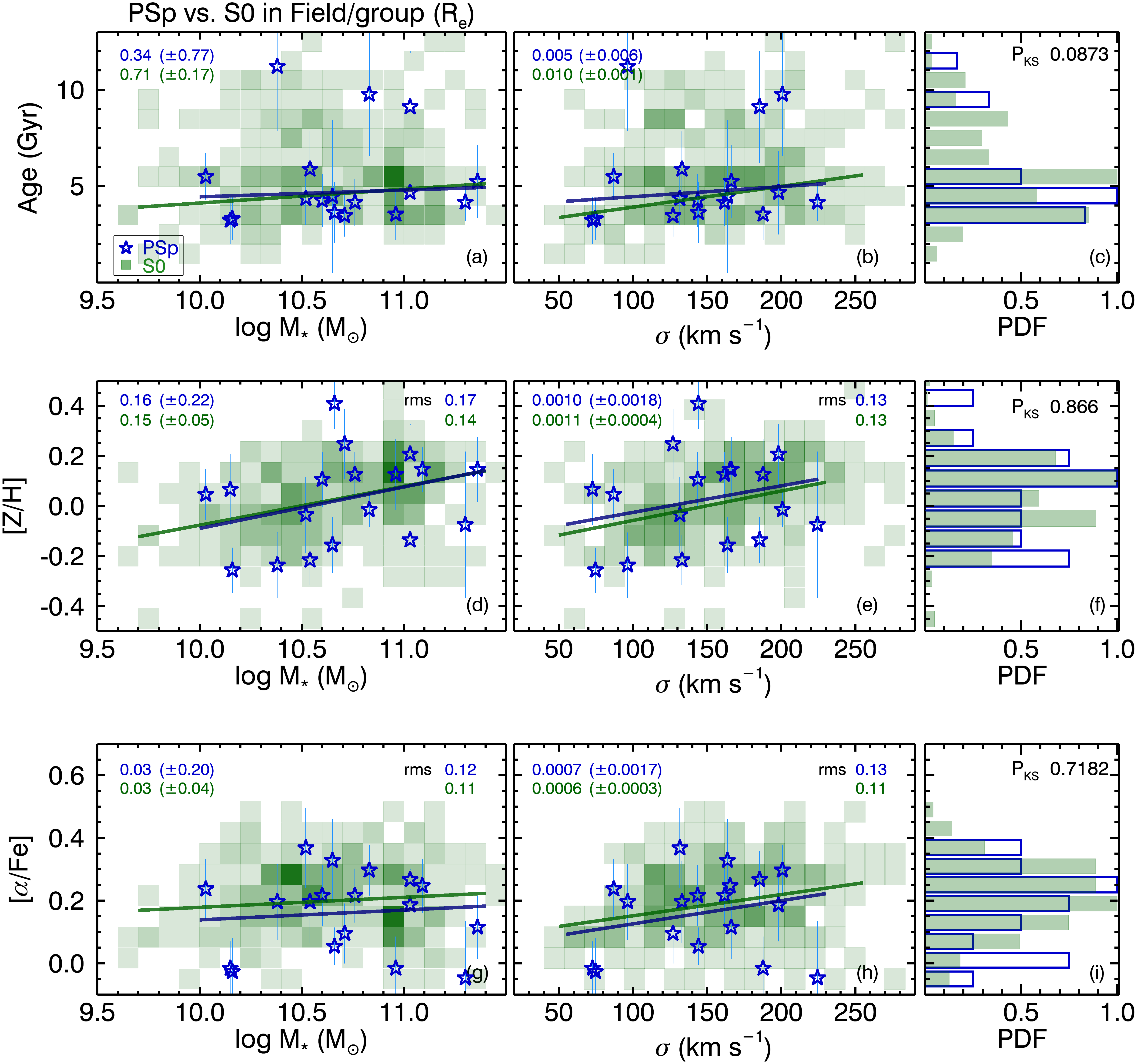}
\caption{The same as Figure \ref{F5} but for the field/group passive spirals and S0s. In histogram, the blue and green bins show passive spirals and S0s, respectively.
\label{F6}}
\end{figure*}
%----------------------------------------

Firstly, the passive spirals are classified by visual inspection (MP and JHL) using composite color images from the fourteenth data release of the Sloan Digital Sky Survey (SDSS DR14; \citealt{Abo18}) and the Dark Energy Camera Legacy Survey (DECaLS; \citealt{Dey16}) for the SAMI sample. We select red spiral galaxies with prominent red arms (e.g. $9016800089$ and $288992$ in Figure \ref{F1}). We also select the galaxies in which the light distribution is not smooth and the arms are open (not a ring). In this case, the arms could be less prominent but still exist (e.g. $9011900418$ and $107528$ in Figure \ref{F1}). The selected passive spirals are classified as from E to Early-spiral classes by \citet{Cor16}. We removed the galaxies that are too inclined to tell if they have spiral arms or not, from the sample: $0\%$ in E ($294$) and E/S0 ($254$) classes, $\sim 25.6 \%$ ($91/355$) in S0 class, $\sim 30.7 \%$ ($104/339$) in S0/Early-spiral class, and $\sim 4.9 \% (17/349$) in Early-spiral class. The galaxies showing clear dust lanes and star-forming arms are also excluded: $\sim 13.0 \%$ ($44/339$) and $\sim 11.2 \%$ ($39/349$) in S0/Early-spiral and Early-spiral classes, respectively. In addition, we removed the four candidates (out of $\sim 1500$ galaxies from E to Early-spiral types) that appear distorted probably due to interactions with neighbors, for which the distinction between spiral arms and tidal features is ambiguous.

Secondly, the galaxies are selected with $4.6-12 \mu$m $\leq 2$ and $3.4-4.6 \mu$m $\leq 0.8$ using WISE colors \citep{Jar17}, which is a good indicator for distinguishing quiescent galaxies and excluding optically red galaxies due to dust-obscured star formation (Figure \ref{F2}). With these criteria, we find $70$ passive spirals: $52$ from the SAMI clusters and $18$ in the SAMI-GAMA regions (group/field environments), respectively. Several examples of the final sample are shown in Figure \ref{F1}. In the finally selected passive spirals, most of passive spirals are classified as S0/Early-spiral and Early-spiral classes ($\sim 71 \%$; $50/70$). The rest are classified as E (4), E/S0 (8), S0 (7), and unknown (1) types. We identified weak spiral structures in the four galaxies classified as Es by \citet{Cor16}.

As a comparison sample, we select S0s by adopting the SAMI visual morphology classification (\citealt{Cor16}). In our analysis, the S0 comparison sample consists of E/S0, S0, and S0/Early-spiral subtypes. Our S0s are also controlled by their WISE colors in the same manner as our passive spirals (Figure \ref{F2}), in order to strictly exclude star-forming S0s. Therefore, in the S0/Early-spiral subtype, the galaxies without arms and with no/weak star formation are classified as control S0 sample. Among all S0s, $273$ and $351$ are selected in the cluster and the GAMA regions, respectively. Hereafter, we refer to our sample as the `cluster passive spirals/S0s' and the `field/group passive spirals/S0s'. Figure \ref{F31} presents the stellar mass and redshift distributions of the final sample. As seen in Figure \ref{F32}, which presents specific star formation rate as a function of mass and the WISE [4.6]-[12] color, our final sample is mostly well distinguished from the star-forming late-type galaxies (blue circles). A few passive spirals and S0s above the threshold of quiescent galaxies (dashed line in the top panel of Figure \ref{F32}) do not change our conclusion.

% Cluster S0s:          273
% Field S0s:          351
\vskip 5mm
%%%%%%%%%%%%%%%%%%%%%%%%%%%%%%%%%%%%%%%%%%%%%%%%%%%%
\subsection{Measurements of the Stellar Population Properties}  % 2.2
%%%%%%%%%%%%%%%%%%%%%%%%%%%%%%%%%%%%%%%%%%%%%%%%%%%%

The measurements of stellar populations are obtained from \citet{Sco17}. The measured 20 Lick indices include five Balmer indices (H$\delta_A$, H$\delta_F$, H$\gamma_A$, H$\gamma_F$, H$\beta$), six iron-dominated indices (Fe4383, Fe4531, Fe5015, Fe5270, Fe5335, Fe5406), four molecular indices (CN1, CN2, Mg1, Mg2) and five more indices in the SAMI blue wavelength range (Ca4227, G4300, Ca4455, C4668, Mg$b$). Each index was measured from the observed spectrum after correcting for broadening and emission lines. Luminosity-weighted simple stellar population-equivalent age, metallicity and $\alpha$-abundance ratio were measured by using the $\chi^2$ minimization method \citep{Pro04} fitted to the stellar population models of \citet{Sch07} and \citet{Tho11}. The full details of the analysis are in section 3 of \citet{Sco18}. We adopt the measurements within a $1.4$ arcsec diameter aperture (center) and within an elliptical aperture with a major axis radius of $R_{\rm e}$. The $1.4$ arcsec diameter mostly corresponds to less than $\sim$half $R_{\rm e}$ for passive spirals. 

%%%%%%%%%%%%%%%%%
\section{Results}  % 3
%%%%%%%%%%%%%%%%%
%---------------------------------------- F7
\begin{figure}
\includegraphics[width=8.7cm]{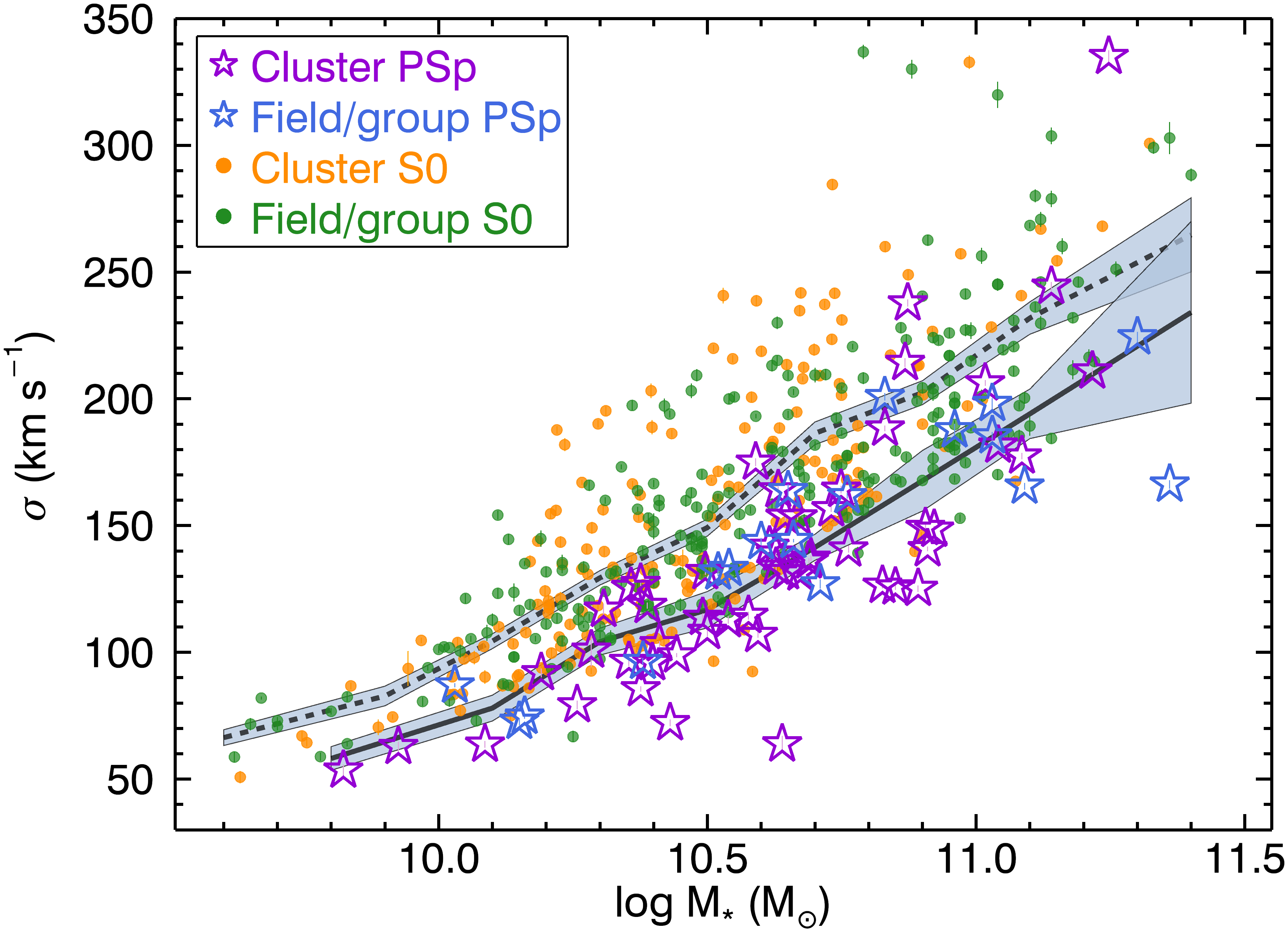}
\caption{Stellar mass vs. $\sigma_{\star}$ for the passive spirals (stars) and the S0s (circles). The mean $\sigma_{\star}$ values per $0.2$ stellar mass bin are shown with solid (cluster and field/group passive spirals) and dashed (cluster and field/group S0s) lines, and shaded areas show the standard error of the mean.
\label{F7}}
\end{figure}
% S0s:   69.0425      98.5588      105.982      137.773      150.731      184.153      200.516      227.677      262.317 
% PSPs:     -         58.1422      78.0080      104.586      118.443      142.886      167.670      194.080      200.497
% delta:    -         40.4167      27.9737      33.1877      32.2883      41.2663      32.8456      33.5968      61.8202
%----------------------------------------- F8
\begin{figure*}
\includegraphics[width=17cm]{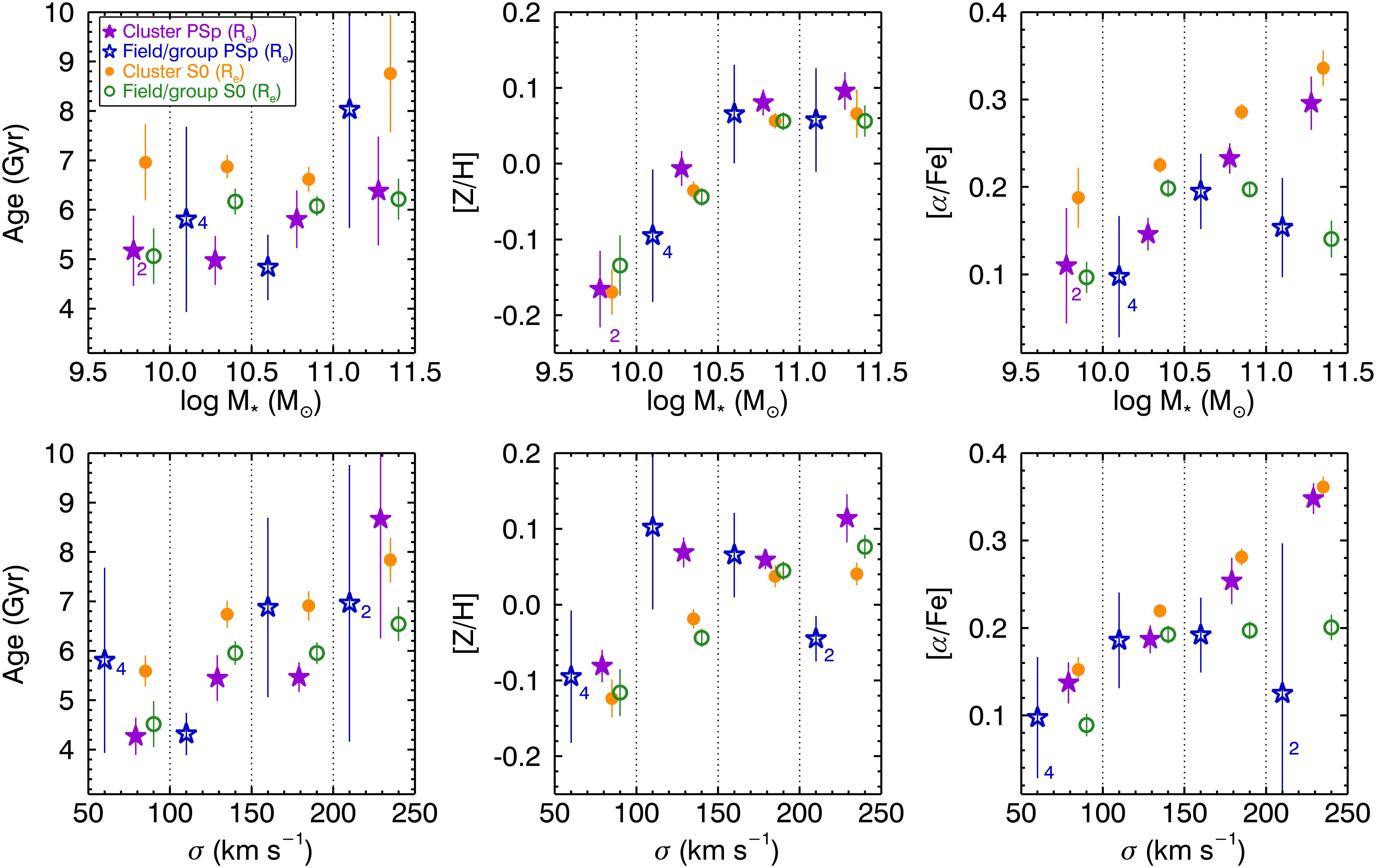}
\caption{Mean age, [Z/H], and [$\alpha$/Fe] in each mass (top panels) and $\sigma_{\star}$ (bottom panels) bin for all subdivisions as shown in the Figures \ref{F4} - \ref{F6}. The error bar is the standard error of the mean. We mark the number of galaxies beside symbol when the galaxies of each bin is $< 5$. The vertical dotted lines identify the width of the bins.
\label{F8}}
\end{figure*}
%----------------------------------------
% 1st. 
%In this section, we compare the stellar populations of passive spirals with those of S0s in different environments.

% 6 PS in isolation
% 5 PS in group with 2 -4 members
% 4 PS in group with 14-20 members
% 3 PS in group with 28 members (same group)
%%%%%%%%%%%%%%%%%%%%%%%%%%%%%%%%%%%%%%%%%%%%%%%%%%%%%%%%%%%%%%%%%%%%%%%%%%
\subsection{Passive spiral galaxies: comparison between cluster and field/group environments}  % 3.1
%%%%%%%%%%%%%%%%%%%%%%%%%%%%%%%%%%%%%%%%%%%%%%%%%%%%%%%%%%%%%%%%%%%%%%%%%%

Figure \ref{F4} presents the age, [Z/H] and [$\alpha$/Fe] within $1.4$ arcsec and $R_{\rm e}$ as a function of stellar mass and velocity dispersion ($\sigma_{\star}$) for the cluster and field/group passive spirals. The $\sigma_{\star}$ is measured from the spectrum inside a circular aperture of radius $R_{\rm e}$, as described in \citet{Sco18} using the method from \citet{Van17}. 

Age shows no correlation with either mass or $\sigma_{\star}$ both in the cluster and field/group passive spirals. [Z/H] shows a very marginal difference between the two environments. In the clusters, [Z/H] tends to increase with mass (and $\sigma_{\star}$) with a large scatter and to be systematically higher in the center ($\alpha \sim 0.22 \pm 0.17$) than within one $R_{\rm e}$ ($\alpha \sim 0.16 \pm 0.16$), whereas such a correlation is hardly detected in the field/groups. The environmental dependence also appears to be tight for [$\alpha$/Fe] ($\alpha \sim 0.17 \pm 0.157$ in the center and $\alpha \sim 0.19 \pm 0.151$ within one $R_{\rm e}$): the cluster passive spirals have obviously higher [$\alpha$/Fe] than the field/group passive spirals at the massive/high-dispersion end.

We investigate the environment of the field/group passive spirals by using the group catalog from \citet{Tre18}. Among $18$ galaxies, six are in isolation, the rest of them are in the groups with a wide range of halo mass (log M$_{\star}$/M$_{\odot}$) $\sim 9.8 - 11.65$. All isolated passive spirals are young ($\lesssim 6$ Gyr) but are not distinct in [Z/H] or [$\alpha$/Fe], as shown in Figure \ref{F4}.

%NGC 4073 group is X-ray bright group \citealt{Jel08}. Number of members are 22 \citealt{Hel00}. 

%%%%%%%%%%%%%%%%%%%%%%%%%%%%%%%%%%%%%%%%%%%%%%%%%%%%%%%%%%%%%%%%%%
\subsection{Passive spirals vs. S0s: comparison between cluster and field/group environments}  % 3.2
%%%%%%%%%%%%%%%%%%%%%%%%%%%%%%%%%%%%%%%%%%%%%%%%%%%%%%%%%%%%%%%%%%

We compare age, [Z/H] and [$\alpha$/Fe] within $R_{\rm e}$ with stellar mass and $\sigma_{\star}$ between passive spirals and S0s in the cluster and field/group. Figure \ref{F5} shows that cluster passive spirals tend to be slightly younger, more metal-rich and less $\alpha$-enhanced than S0s as a function of mass (and $\sigma_{\star}$), which results in the clear differences of the histograms. A two-dimensional Kolmogorov-Smirnov (KS) test of similarity in the cumulative distributions between passive spirals and S0s yields a probability (P$_{KS}$) of $0.0009$ for age, $0.028$ for [Z/H] and $0.0001$ for [$\alpha$/Fe]. These differences are clearer in the lower mass (or $\sigma_{\star}$) range. 

Unlike in the clusters, the distribution of [$\alpha$/Fe] shows no notable difference (P$_{KS} \sim 0.7$ in Figure \ref{F6}i) between the field/group passive spirals and S0s in Figure \ref{F6}. Interestingly, the [$\alpha$/Fe] - mass($\sigma_{\star}$) correlation slope in the field/groups appears to be shallower than in the clusters for S0, as well as for passive spirals. The slope in [$\alpha$/Fe] is nearly constant ($\alpha \sim 0.03\pm0.20$) or even decreasing with stellar mass at the high-mass end for the field/group S0s, similar to the passive spirals ($\alpha \sim 0.03\pm0.04$). In addition, field/group S0s tend to have slightly larger scatter in both [Z/H] and [$\alpha$/Fe] compared to cluster S0s. This is particularly evident at the high-mass end (see rms values in the top right corner of the Figures \ref{F5} and \ref{F6}). We estimated the spearman's rank coefficient for [$\alpha$/Fe], to quantitatively compare its mass ($\sigma_{\star}$) dependence between environments: $0.56$ ($0.63$) for cluster PSp and $-0.06$ ($0.09$) for field/group PSp, and $0.36$ ($0.65$) for cluster S0s and $0.04$ ($0.24$) for field/group S0s. Since the comparisons show that massive field/group PSp and S0 galaxies tend to have lower [$\alpha$/Fe] than cluster ones, we suspect that galaxies in low-density environments may have more extended star-formation histories \citep{deL11}. The results of spearman's rank coefficient and the significance for [Z/H] and [$\alpha$/Fe] are summarized in Table \ref{Tab1}.

%--------------------------------------------------------------------
\begin{deluxetable}{lllcc}
\tablenum{1} \tablecolumns{5} \tablecaption{The spearman's rank correlation coefficient ($\rho$) and significance of deviation (P$_{0}$) for [Z/H] and [$\alpha$/Fe] with stellar mass and $\sigma$} \tablewidth{0pt}
\tablehead{   & & & PSp & S0 \\
              & & & $\rho$ (P$_{0}$) & $\rho$ (P$_{0}$) }
\startdata
[Z/H] & Cluster     & log M$_\star$ (M$_\odot)$  & 0.53 (5e-05) & 0.48 (4e-17) \\
      &             & $\sigma$ (km s$^{-1}$)     & 0.45 (0.001) & 0.41 (2e-12) \\
      & Field/group & log (M$_\star$/M$_\odot)$  & 0.42 (0.080) & 0.38 (5e-13) \\
      &             & $\sigma$  (km s$^{-1}$)    & 0.22 (0.367) & 0.40 (3e-14) \\
\hline 
[$\alpha$/Fe] & Cluster     & log (M$_\star$/M$_\odot)$  & 0.56 (1e-05) & 0.36 (9e-10) \\
              &             & $\sigma$  (km s$^{-1}$)    & 0.63 (6e-07) & 0.65 (2e-33) \\
              & Field/group & log (M$_\star$/M$_\odot)$  & -0.06 (0.823) & 0.04 (0.492) \\
              &             & $\sigma$  (km s$^{-1}$)    &  0.09 (0.726) & 0.24 (9e-06) 
\enddata
\tablecomments{The Spearman's rank correlation coefficient ($\rho$) will be between a value of $-1$ and $+1$. $\rho = -1$ indicates a perfect negative correlation and $\rho = +1$ indicates a perfect positive correlation.}
\label{Tab1}
\end{deluxetable}
%--------------------------------------------------------------------

Figure \ref{F7} shows the stellar mass versus $\sigma_{\star}$ for the passive spirals and S0s. The $\sigma_{\star}$ of passive spirals tends to be lower (mean $\Delta \sigma$ in all stellar mass range \textbf{$\sim 32$} km s$^{-1}$) than that of S0s at a given stellar mass. This may imply that some S0s suffer a kind of dynamical heating after star formation stops. Loss of stellar mass in the galaxies could be also playing a role in mass$- \sigma$ relations \cite{Jos20}. The environmental dependence of the stellar mass$- \sigma$ relations at a given galaxy type appears not to be significant.

In Figure \ref{F8}, in order to compare all together, we show the mean of stellar population parameters with stellar mass (top panels) and $\sigma_{\star}$ (bottom panels) bins for all subdivisions as shown by the Figures \ref{F4} - \ref{F6}. We present the number of galaxies in case the galaxies less than $5$ in each bin for caution. It is clear that [$\alpha$/Fe] flattens and/or bends over at log (M$_\star$/M$_\odot$) $\gtrsim 10.5$ or $\sigma \gtrsim 100$ km s$^{-1}$ for the field/group passive spirals and S0s.

%----------------------------------------- F9
\begin{figure*}
\includegraphics[width=18cm]{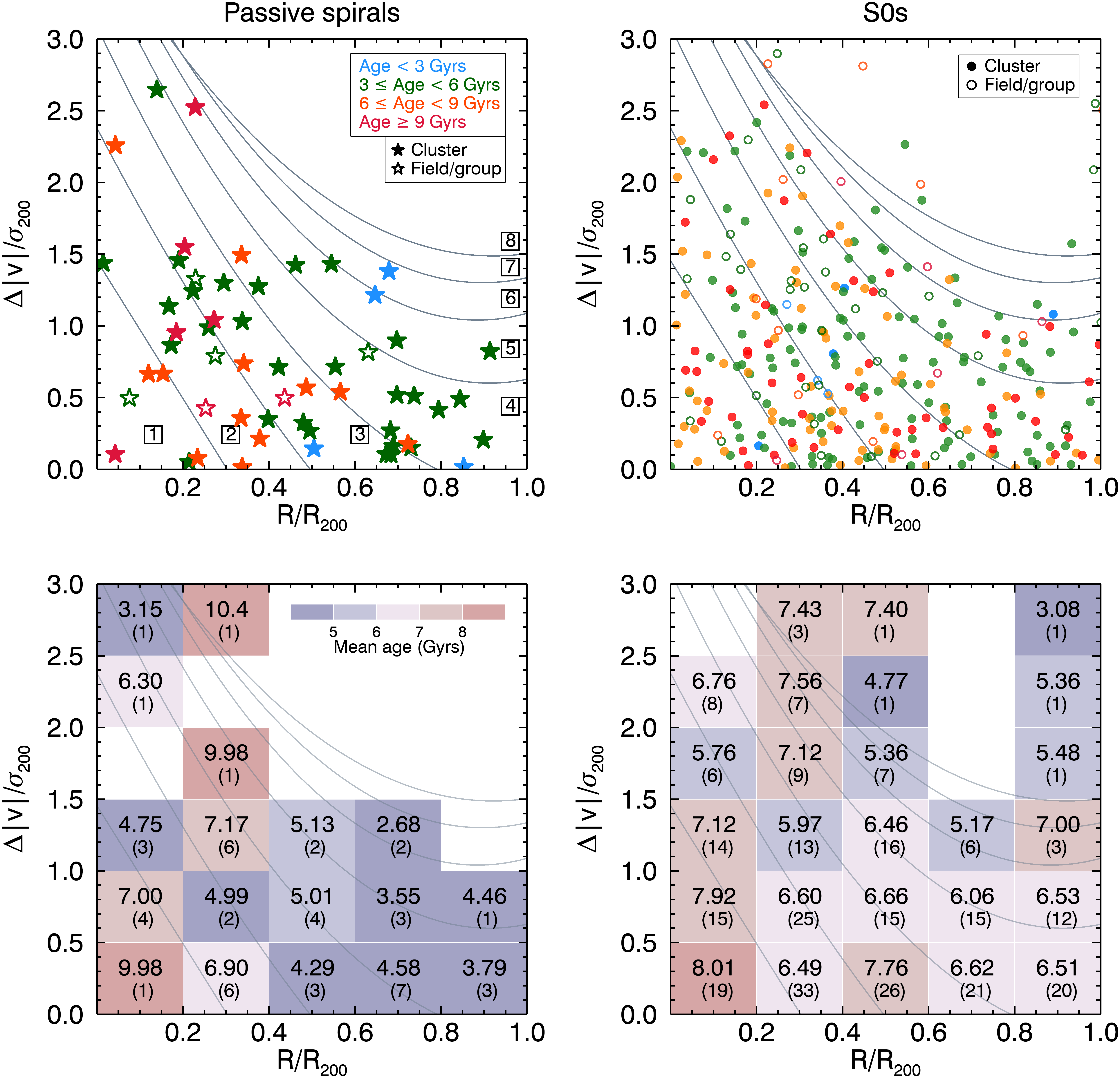}
\caption{Top row: The age distribution of passive spirals (left column) and S0s (right column) in cluster and field/group environments across projected phase-space. Red: older than $9$ Gyrs, Orange: $6 \leq$ Age $< 9$ Gyrs, Green: $3 \leq$ Age $< 6$ Gyrs, Blue: younger than $3$ Gyrs. The $\sigma_{200}$, and R$_{200}$ of eight clusters are taken from table 1 in \citet{Owe17}. The solid lines show the zones from 1 to 8 defined in \citet{Pas19}. Bottom row: the mean age and the number of galaxies in each bin are presented. 
\label{F9}}
\end{figure*}
%-----------------------------------------
%----------------------------------------- F10
\begin{figure*}
\includegraphics[width=18cm]{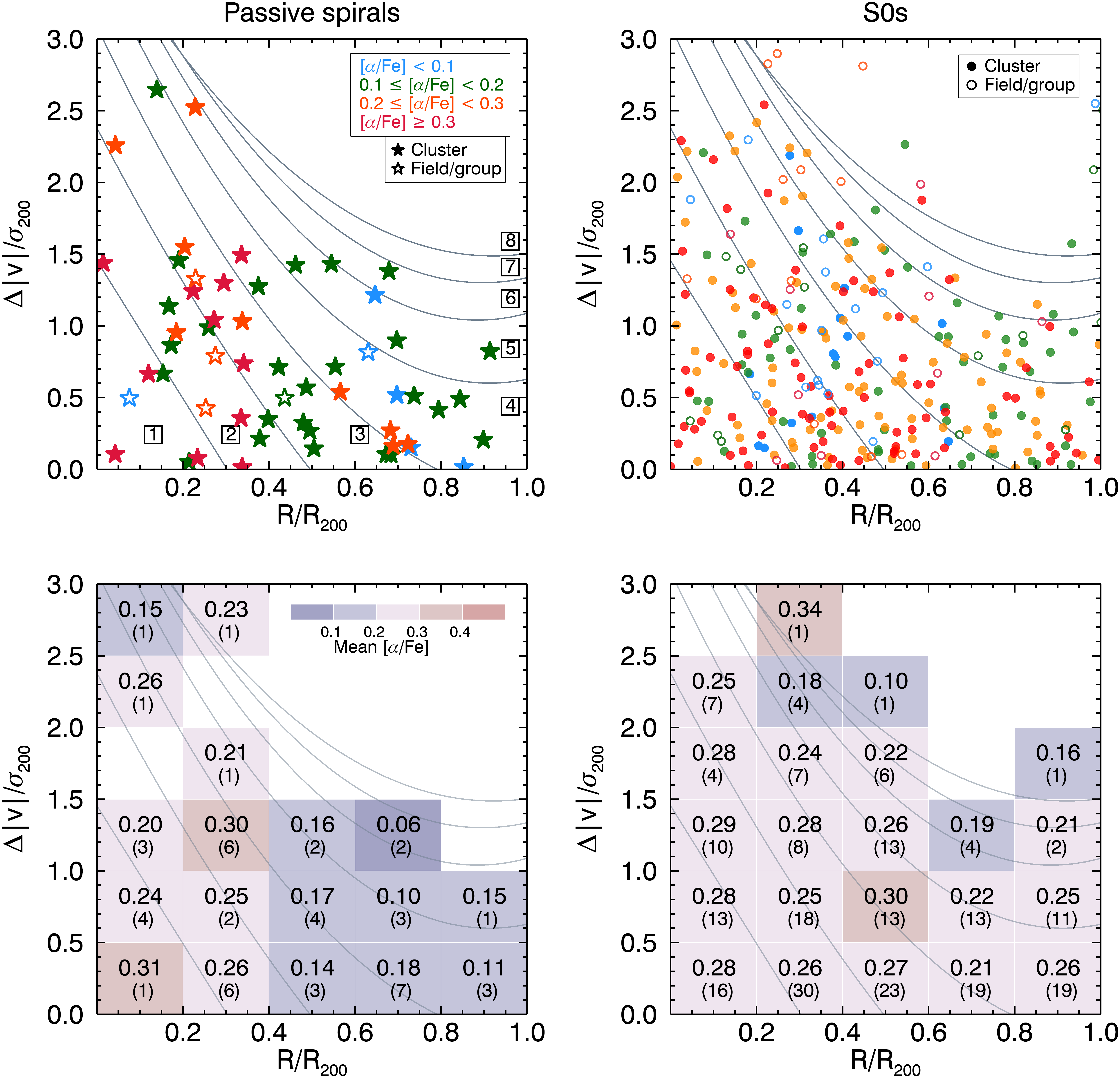}
\caption{Top row: The [$\alpha$/Fe] distribution of passive spirals (left column) and S0s (right column) in cluster and field/group environments across projected phase-space. Red: [$\alpha$/Fe] $\geq 0.3$, Orange: $0.2 \geq$ [$\alpha$/Fe] $< 0.3$, Green: $0.1 \geq$ [$\alpha$/Fe] $< 0.2$ Gyrs, Blue: [$\alpha$/Fe] $< 0.1$. The $\sigma_{200}$, and R$_{200}$ of eight clusters are taken from table 1 in \citet{Owe17}. The solid lines show the zones from 1 to 8 defined in \citet{Pas19}. Bottom row: the mean [$\alpha$/Fe] and the number of galaxies in each bin are presented.
\label{F10}}
\end{figure*}
%-----------------------------------------
%----------------------------------------- F11
\begin{figure*}
\includegraphics[width=18cm]{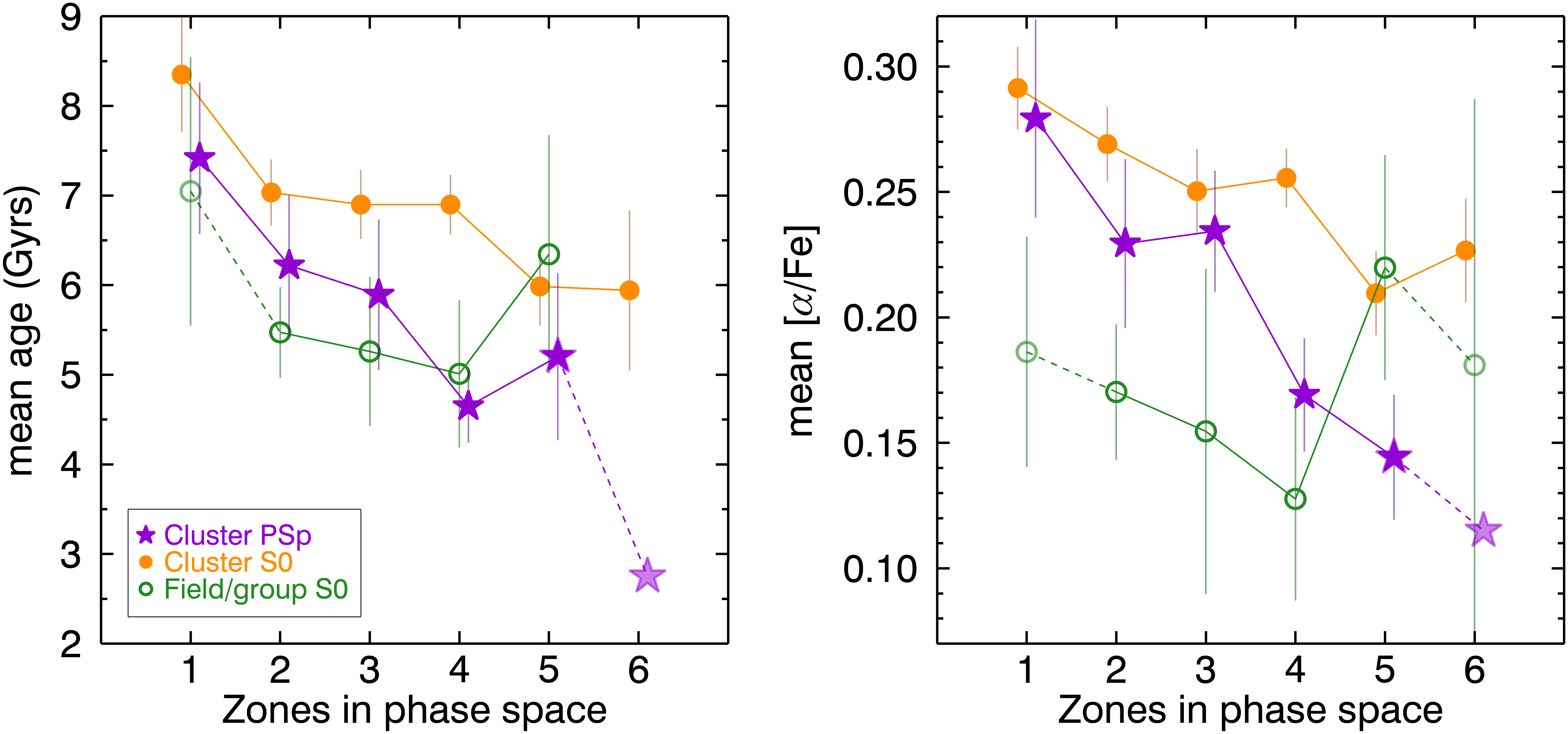}
\caption{Mean age and [$\alpha$/Fe] as a function of zone in projected phase-space presented by Figures \ref{F9} and \ref{F10}. Stars are cluster passive spirals. Field/group passive spirals are not presented due to the small sample size (four galaxies). Orange filled and green open circles are cluster and field/group S0s, respectively. The error bar is the standard error of the mean. We connect to the symbols with dashed lines when the galaxies of each zone is less than $5$. 
\label{F11}}
\end{figure*}
%-----------------------------------------

%%%%%%%%%%%%%%%%%%%%
\section{Discussion}
%%%%%%%%%%%%%%%%%%%%
In the previous section we have shown that stellar populations show different trends between cluster and field/group passive spirals, especially for [$\alpha$/Fe]. Here, we discuss the possible origins of these different trends between environments and between passive spirals and S0s. 

From our results, [$\alpha$/Fe] increases with mass (and $\sigma_\star$) for both passive spirals and S0s, in the clusters. This finding is consistent with previous studies: for early-type galaxies both metallicity and $\alpha$-abundance correlate tightly with stellar mass (and $\sigma_\star$; \citealt{Ber06}; \citealt{McD15}). In particular, the [$\alpha$/Fe] - mass correlation (Figures \ref{F5}g, \ref{F5}h) is an evidence that the more massive galaxies quenched in a short timescale at earlier epochs. Notably, cluster passive spirals have slightly younger age and, crucially, lower $\alpha$-abundance compared to cluster S0s (Figure \ref{F5}). If passive spirals and S0s commonly originate from star-forming spiral galaxies, this means that passive spirals had undergone star formation for a longer timescale and became quiescent more recently than S0s. The mean age of cluster and field/group passive spirals is $\sim 5$ Gyrs. This is coincident with the simulations from \citet{Bek02}, in which spiral arm structures are found to fade over several Gyrs, after gas is stripped.

\citet{Bek11} argued that the progenitors of cluster S0s might be reshaped and quenched even before cluster infall, through mergers and/or tidal interactions in galaxy groups or cluster outskirts (pre-processing; \citealt{DeL12}, \citealt{Oh18}). On the other hand, the fact that passive spirals are younger and less $\alpha$-enhanced than S0s in the clusters suggests that passive spirals may have reached the clusters without severe interactions with neighbors. After infall, they may have experienced star formation quenching by cluster environmental effects such as thermal evaporation (\citealt{Cow77}), ram-pressure stripping (\citealt{Gun72}), starvation or strangulation (\citealt{Lar80}; \citealt{Bek02}). This would naturally explain why passive spirals show younger age and longer star-formation timescale than S0s.

As an additional check for the scenario for clusters, we inspect the distribution of stellar ages and [$\alpha$/Fe] of all the cluster passive spirals in the projected phase-space diagram (Figures \ref{F9} and \ref{F10}). We compare the observed distribution with the analytic curves from \citet{Pas19}, which grade the average infall time ($=$ time elapsed since the first infall) of the galaxies. The distribution of the average infall time of the galaxies systematically shifts from high to low value with increasing zone number in their simulations (1 for the earliest infall, while 8 for the latest infall; see figure 4 in \citealt{Pas19}). The stellar ages and [$\alpha$/Fe] of passive spirals and S0s show an overall shift from older to younger ages and from more $\alpha$-enhanced to less $\alpha$-enhanced along R/R$_{200}$ and $\Delta v$/$\sigma_{200}$: galaxies with smaller R/R$_{200}$ and $\Delta v$/$\sigma_{200}$ tend to have suffered quenching earlier. This is clearly seen in Figure \ref{F11}. Furthermore, cluster S0s show older mean age and more $\alpha$-enhanced than cluster passive spirals in all zones.  This is similar to the results from \citet{Pas19} and \citet{Smi19} with much larger SDSS sample. 

In summary, star-forming spiral galaxies may be quenched through mergers or interactions before falling into a cluster, becoming relatively old and $\alpha$-enhanced S0s. Otherwise, without significant tidal events, spiral galaxies may arrive in the infall region of a cluster with ongoing star formation. Such spiral galaxies may suffer star formation quenching by cluster environmental effects and eventually evolve into S0s via the transitional phase of passive spirals. Those S0s (or passive spirals) have younger ages and lower $\alpha$-enhancements than the S0s formed through mergers at an earlier epoch.

%% GAMA
As shown in Figure \ref{F8}, S0s tend to be more $\alpha$-enhanced (by $0.06 - 0.08$ dex at a given mass bin) than passive spirals in the cluster environments, while their metallicities are similar, which indicates that cluster S0s have suffered more rapid chemical enrichment and star formation quenching than cluster passive spirals. However, such a difference in [$\alpha$/Fe] between passive spirals and S0s in clusters is hardly found in massive field/group passive spirals and S0s, which implies that their formation processes in low-density environments may be different from those in clusters.

In field/group environments, the trend of [$\alpha$/Fe] is different from the well-known linear correlation between [$\alpha$/Fe] and stellar mass: they appear to be flattened or even decreasing with increasing mass at log (M$_{\star}$/M$_{\odot}) \gtrsim 10.5$, both for passive spirals and S0s (see Figure \ref{F6}g and h). According to \citet{Ber06}, the stars in early-type galaxies formed at slightly earlier times and on a slightly shorter timescale in dense regions than in less dense regions. Our finding shows that the different quenching timescale between high- and low-density environments is due to the difference at the high-mass end. At log (M$_{\star}$/M$_{\odot}) \gtrsim 10.5$, passive spirals or S0s in low-density environments appear to have longer star-formation timescales, unlike those in high-density environments. That is, in low-density environments, star-formation quenching mechanisms may be weak or, alternatively, there may be some mechanisms that prolong star formation.

One possible scenario for low-density environments is that those massive galaxies may be constantly rejuvenated by interactions with gas-rich neighbor galaxies, which is known to be more frequent in low-density environments with relatively low velocity-dispersion (\citealt{Too72}, \citealt{Lav88}, \citealt{Byr90}). Radial transfer of gas to the central region and subsequent triggering of new star formation may influence light-weighted [Z/H] and [$\alpha$/Fe] in interacting galaxies. Although the change of metallicity by interactions may depend on the properties of the interacting neighbors, [$\alpha$/Fe] is expected to decrease as star formation is going on. Alternatively, massive passive spirals and S0s in low-density environments may be supplied with fuel for star formation by tidal stirring \citep{May01} or gas accretion. Since gas stripping is not strong in low-density environments, they can keep forming stars using accreted gas, and consequently become relatively young and $\alpha$-depressed galaxies. However, the suggested scenarios need to be verified using a much larger sample, because the sample size of our low-density galaxies ($18$) is not sufficient.

%Therefore, we also suggest that this is a direct evidence that environments may play key role in, at least, star formation quenching of galaxies.
%----------------
%Figure \ref{F43} presents the spatial distributions of eight clusters with ordered in increasing mass of clusters and marked the location of passive spirals (stars) and members (dots) observed in SAMI survey in each cluster. In terms of the cluster environments, passive spirals are located in everywhere from the center of the cluster to the outskirts. 
%\begin{figure*}
%\includegraphics[width=18cm]{F4_3.png}
%\caption{Spatial distribution of \textcolor{violet}{57} cluster passive spirals. Stars and circles are passive spirals and member galaxies in each cluster, respectively. (Cluster information is taken from Table 1 of \citet{Owe17}).
%\label{F43}}
%\end{figure*}
%----------------
\vskip 1.5mm

%%%%%%%%%%%%%%%%%%%%%%%%%%%%%%%%%
\section{Summary and Conclusions}
%%%%%%%%%%%%%%%%%%%%%%%%%%%%%%%%%
We have investigated the stellar populations of passive spirals with respect to their environment using a large sample from the SAMI Galaxy Survey. We found that star formation timescale of passive spirals appears to be different between clusters and field/groups. In the low-density environments, passive spirals show no [$\alpha$/Fe]-mass correlation, and have low [$\alpha$/Fe] values even at the highest stellar mass. This indicates that passive spiral galaxies may have different formation histories depending on their environments, which confirms the previous findings that environmental effects play a key role on the star-formation quenching in spiral galaxies (\citealt{Bar18}; \citealt{Sch17}; \citealt{Sch19}).

We discussed possible processes that can result in the observed trends, focusing on the difference between the environments. Like massive galaxies in clusters, those in low-density regions may have experienced burst of star formation making them old, metal-rich and $\alpha$-enhanced at earlier epochs as well. However, cold gas accretion may occur more easily in low-density environments, which consequently prolongs star formation - with the effect being most noticeable at the high-mass end. Alternatively, gas fuel by wet minor mergers can also make lower [$\alpha$/Fe]. 

In clusters, there may be various and complex mechanisms that accelerate galaxy evolution. However, here we suggest two simplified channels that transform spirals to S0s through the passive spiral phase (\citealt{Owe19}): (1) S0s may be formed by late-late or early-late type merger at early epoch before infall to a cluster halo. The starburst triggered by merger and subsequent mass-quenching may have made them old, metal-rich and $\alpha$-enhanced. (2) Spirals may be transformed into S0s via the passive spiral phase, by losing gas without morphological distortion. Their gas can be stripped away predominantly in the halo and outer disks by ram-pressure stripping or other cluster mechanisms, while passing through the cluster. After losing their gas, spirals become quiescent, and evolve secularly holding their spiral structures a few Gyrs - this is the passive spiral phase \citep{Bek02}. Our findings provide observational evidence that there are multiple channels to form passive spirals and the formation mechanisms closely depend on their environments. Future studies on the detailed structures and kinematics of passive spirals may be beneficial to verify and elaborate our scenarios.

%%%%%%%%%%%%%%%%%%%%%%%%%%%%%%%%%
\section*{Acknowledgments}

We gratefully thank the anonymous referee for constructive comments that have significantly improved this manuscript. This work was supported by UST Young Scientist Research Program through the University of Science and Technology (No. 2019YS10). The SAMI Galaxy Survey is based on observations made at the Anglo-Australian Telescope. The Sydney-AAO Multi-object Integral field spectrograph (SAMI) was developed jointly by the University of Sydney and the Australian Astronomical Observatory. The SAMI input catalogue is based on data taken from the Sloan Digital Sky Survey, the GAMA Survey and the VST ATLAS Survey. The SAMI Galaxy Survey is supported by the Australian Research Council Centre of Excellence for All Sky Astrophysics in 3 Dimensions (ASTRO 3D), through project number CE170100013, the Australian Research Council Centre of Excellence for All-sky Astrophysics (CAASTRO), through project number CE110001020, and other participating institutions. The SAMI Galaxy Survey website is \href{url}{http://sami-survey.org/}. Parts of this research were conducted by the Australian Research Council Centre of Excellence for All Sky Astrophysics in 3 Dimensions (ASTRO 3D), through project number CE170100013. JvdS acknowledges support of an Australian Research Council Discovery Early Career Research Award (project number DE200100461) funded by the Australian Government. FDE acknowledges funding through the H2020 ERC Consolidator Grant 683184. JBH is supported by an ARC Laureate Fellowship that funds Jesse van de Sande and an ARC Federation Fellowship that funded the SAMI prototype. H.J. is supported by the National Research Foundation of Korea(NRF) grant funded by the Korea government(MSIT) (No. 2019R1F1A1041086). TMB is supported by an Australian Government Research Training Program Scholarship. JJB acknowledges support of an Australian Research Council Future Fellowship (FT180100231). MSO acknowledges the funding support from the Australian Research Council through a Future Fellowship (FT140100255).

\bibliographystyle{apj}
%\bibliography{ref.bib}  

\end{document}